\documentclass[11pt,twoside]{article}

\usepackage{asp2006}
\usepackage{epsf}
\usepackage{psfig}
\usepackage{lscape}
\usepackage{graphicx}

\begin{document}
\title{Zeeman Component Decomposition (ZCD)\\ of polarized spectra:
Application for the quiet Sun internetwork magnetic field}   
\author{C.~Sennhauser, \altaffilmark{1}, S.V.~Berdyugina \altaffilmark{2}}   
\altaffiltext{1}{Institute for Astronomy, ETH Zurich, 8093 Zurich, Switzerland}
\altaffiltext{2}{Kiepenheuer-Institut f\"ur Sonnenphysik, 79104 Freiburg, Germany}


\begin{abstract} 
Multiline techniques assuming similar line profiles have become a standard tool in stellar astronomy for increasing the signal-to-noise ratio (SNR) of spectropolarimetric measurements. However, due to the widely-used weak field approximation their benefits could not so far be used for solar observations, where a large variety of Stokes profiles emerge from local magnetic fields and measuring weak fields in the quiet Sun remains a challenge. The method presented here permits us to analyze many lines with arbitrary Zeeman splitting and to simultaneously deploy Stokes $IQUV$ spectra to determine a common line profile with the SNR increased by orders of magnitude. The latter provides a valuable constraint for determining separately field strengths for each contributing absorber. This method represents an extension of our recently developed technique of Nonlinear Deconvolution with Deblending  \citep[NDD,][]{sennhauseretal2009}, which accounts for the nonlinearity in blended profiles. Equipped with all those abilities, ZCD is the perfect tool to further increase the informative value of high-precision polarimetric observations.
\end{abstract}


\section{Introduction}

Magnetic fields in the quiet Sun being ubiquatuous but extremely weak and mixed in polarities remain a challenge to detect and study. They have drawn recently more attention, first, because they are thought to play an important role in solar dynamo and heating the chromosphere and, second, because the Sun is so unusually quiet during the current minimum. A coherent picture is however still missing. For instance, measurements of the internetwork magnetic field strength by different methods differ by an order of magnitude, from roughly 10$\,$G up to several hundred G (\citeauthor{litesetal2008} \citeyear{litesetal2008}, \citeauthor{suarezetal2007} \citeyear{suarezetal2007}). This is obviously due to the limited accuracy and/or spatial resolution of current spectropolarimetric observations, which at best achieve the noise level of $10^{-3}$ at the resolution of 0.3 arcsec, as with the SP instrument on Hinode. Here we propose a method which can boost the signal-to-noise ratio (SNR) of spectropolarimetric observations by orders of magnitude and, thus, is capable of providing a more sensitive constraint on very weak internetwork magnetic fields. Our method is based on a new technique called Nonlinear Deconvolution with Deblending \citep[NDD,][]{sennhauseretal2009}, which extracts a common Stokes profile from many spectral lines while accounting for the nonlinearity in blends.


\section{Zeeman Component Decomposition}

Using the advantages of the NDD, we have developed a new method called Zeeman Component Decomposition (ZCD), which is a multiline technique for extracting common Stokes line profiles from spectropolarimetric observations \citep{sennhauserberdyuginainprep}. Our aim is to overcome the limitations imposed by the widely-used weak field approximation and, thus, be able to combine spectral lines with arbitrary splitting patterns, irrespective of the magnetic field strength and complexity in the line forming region. In this case, the common line profile to be retrieved becomes independent of $B$, which is then a parameter we evaluate. All other physical conditions of the stellar (solar) atmosphere in which the lines are formed remain imprinted in the shape of the common line profile, which represents a sum of contributions from various sources in the atmosphere. The initial condition for obtaining such a sum is that the source function $S_i\left(\tau\right)$ is a linear function of optical depth $\tau$ in the formation region of each spectral line $i$. Note that the only requirement is a linearity, while actual parameters defining the precise behaviour of $S_i$ may vary for different absorption lines. To combine individual profiles from lines with multiple Zeeman components (anomalous Zeeman effect), we further assume that the shapes of subcomponents are equal, following the hypothesis of complete redistribution (e.g. \citeauthor{landi1976} \citeyear{landi1976}).

To summarize, the ZCD comprises the following features:
\begin{itemize}
\item Accurate deconvolution of spectra to disentangle nonlinear contributions from different lines to blends.
\item Ability to deal with arbitrary Zeeman multiplets, in particular when
Zeeman splitting becomes larger than Doppler broadening.
\item Account for individual saturation levels of Zeeman components leading to different profile shapes.
\item Precise treatment of nonlinear blending of $\sigma_\pm$ and $\pi$ components under the assumption of a Milne-Eddington atmosphere.
\item Simultaneous processing of Stokes $I$, $Q$, $U$, and $V$, resulting in a significant additional constraint for retrieving a common Zeeman component profile.
\item No confinement to weak magnetic fields.
\end{itemize}


\section{Results}

\begin{center}
  \begin{figure}[!h]
	\plotone{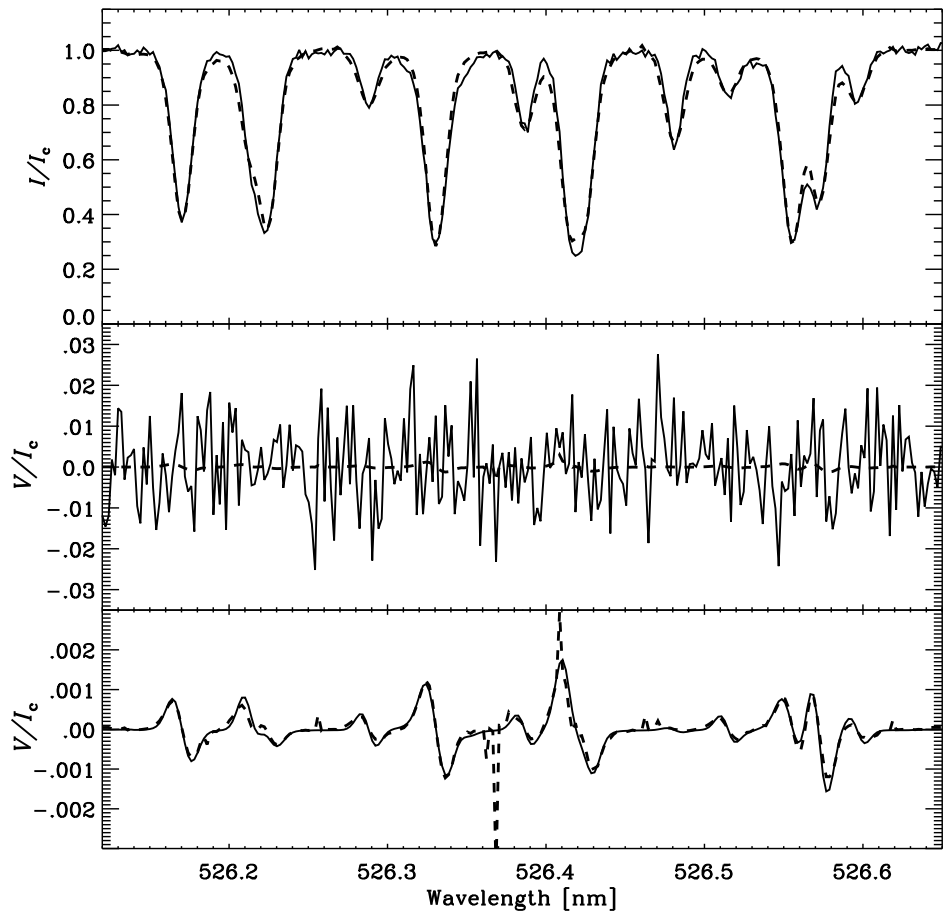}
	\caption{\textit{Top and middle panel:} A section of a simulated spectrum (B=10$\,$G, 1\% noise level, solid lines) of Stokes $I$ and $V$, and  fits obtained with the ZCD using the extracted common line profile (dashed lines).$\quad\quad$ \textit{Lower panel:} Stokes $V$ fit (dashed line) from the middle panel on an axis of ordinates with higher resolution. For comparison, the original Stokes $V$ without noise (thin solid line) is overplotted.}
	\label{fig:fits}
	\end{figure}
\end{center}

We illustrate the capability of our method to extract Stokes profiles by using simulated data. We synthesize a solar-type spectrum in the optical wavelength region 521.5-529.8 nm with 35 lines included in the list at 
a magnetic field of 2, 10, 100, and 500$\,$G directed toward an observer ($\gamma=0$) and the noise levels of $10^{-3}$  and $10^{-2}$ (see Fig.~\ref{fig:fits}). Magnetooptical effects are neglected for the moment. When extracting a common line profile from such spectra we achieve the SNR of $\sim$ 20'000 and 2000 and can confidently detect polarimetric amplitudes of $\sim$0.005\% and 0.05\%, for the two noise levels respectively. These correspond to a standard error of 6$\,$G and 0.85$\,$ G, respectively, in the modulus of B. The errors can be further improved if a wider spectral range with more contributing lines is analyzed.

Employing our technique for measuring weak magnetic fields in the quiet Sun can significantly improve their detection level with the Zeeman effect. This combined with the Hanle effect diagnostics \citep[e.g.,][]{kleintetal2010} can finally reveal the morphology of the solar internetwork magnetic field.


Considering the extremely small amplitudes of the extracted Stokes $V$ signal (Fig.~\ref{fig:fits}, middle and lower panel), it is clear that it was completely embedded in the noise in the original spectrum. We conclude therefore that our method can successfully detect very weak solar magnetic fields even in noisy measurements (obtained, for instance, with high spatial and temporal resolution) if a wide range of spectral information is available. The latter can be achieve for instance by using echelle or FTS spectrographs combined with imaging polarimetry.

Note that possible unknown blends, as included in the simulated spectrum, do not drastically affect the functionality of our code. Indeed, when including very weak lines, noise from the local line profile will be multiplied by a large number, which in some cases turns out to be worse than simply omitting the line from the procedure.


\section{Conclusions and outlook}

We have demonstrated by finding the common line pattern for Stokes $I$ and $V$ simultaneously, that our ZCD code is capable of identifying Zeeman signatures which are far below the noise level of the original spectrum. It directly returns reliable magnetic field strengths, as well as a common line profile containing the physical conditions of the line forming atmosphere.

A full implementation of the method \citep{sennhauserberdyuginainprep} enables the ZCD to recover a magentic field \emph{vector}, i.e., strength and orientation, by inverting a full set of Stokes parameters.  There, magnetooptical effects are included as well, in order to take into account linear polarization signals originating from anomalous dispersion. In addition, our method can be successfully applied to molecular bands despite severe blending of many lines \citep{sennhauseretal2009}. Furthermore, an extension of the ZCD for the (partial) Paschen-Back regime enables investigation of multi-Tesla magnetic objects using both atomic and molecular lines. While being excellent tracers of magnetic activity in cooler regions, molecular lines demonstrate departures from the Zeeman regime at relatively low magnetic field strengths, $\sim 100$ G (\citeauthor{berdyuginaetal2005} \citeyear{berdyuginaetal2005}) which makes it crucial for them to be treated in the Paschen-Back regime.


\acknowledgements This work is supported by the SNF grant PE002-104552. SVB acknowledges the EURYI Award from the ESF (www.est.org/euryi).
\\


\end{document}